\documentclass[10pt,journal,compsoc]{IEEEtran}

\newcommand{\cmark}{\ding{51}}%
\newcommand{\xmark}{\ding{55}}%

\usepackage{forest}
\usepackage{graphicx}
\usepackage{amsmath}
\usepackage{amssymb}
\usepackage{booktabs}
\usepackage[utf8]{inputenc}

\usepackage{algorithm}
\usepackage{algpseudocode}
\usepackage{pifont}
\usepackage{listings}
\usepackage{etoolbox}
\usepackage{lipsum}
\usepackage{mathtools}

\definecolor{commentgreen}{rgb}{0.0, 0.27, 0.13}

\usepackage[pagebackref,breaklinks,colorlinks]{hyperref}

\usepackage[capitalize]{cleveref}
\crefname{section}{Sec.}{Secs.}
\Crefname{section}{Section}{Sections}
\Crefname{table}{Table}{Tables}
\crefname{table}{Tab.}{Tabs.}

\begin{document}

\title{SPCXR: Self-supervised Pretraining using Chest X-rays Towards a Domain Specific Foundation Model}

\author{Syed Muhammad Anwar$^1$ \vspace{0.1cm}
\qquad
Abhijeet Parida$^1$
\qquad
Sara Atito$^2$
\qquad
Muhammad Awais$^2$ \\
\qquad
Gustavo Nino$^1$
\qquad
Josef Kitler$^2$
\qquad
Marius George Linguraru$^1$ \vspace{0.1cm} \\
\qquad
$^1$Children's National Hospital, Washington DC, USA\\
$^1$ {\tt\small \{sanwar,aparida,gnino,mlingura\}@childrensnational.org}\\
$^2$ CVSSP, University of Surrey, Surrey, UK\\
$^2$ {\tt\small \{sara.atito,muhammad.awais,j.kittler\}@surrey.ac.uk}}
\maketitle

\begin{abstract}
   Chest X-rays (CXRs) are a widely used imaging modality for the diagnosis and prognosis of lung disease. The image analysis tasks vary. Examples include pathology detection and lung segmentation. There is a large body of work where machine learning algorithms are developed for specific tasks. A significant recent example is Coronavirus disease (covid-19) detection using CXR data. However, the traditional diagnostic tool design methods based on supervised learning are burdened by the need to provide training data annotation, which should be of good quality for better clinical outcomes. Here, we propose an alternative solution, a new self-supervised paradigm, where a general representation from CXRs is learned using a group-masked self-supervised framework. The pre-trained model is then fine-tuned for domain-specific tasks such as covid-19, pneumonia detection, and general health screening. We show that the same pre-training can be used for the lung segmentation task. Our proposed paradigm shows robust performance in multiple downstream tasks which demonstrates the success of the pre-training. Moreover, the performance of the pre-trained models on data with significant drift during test time proves the learning of a better generic representation. The methods are further validated by covid-19 detection in a unique small-scale pediatric data set. The performance gain in accuracy ($\sim25\%$) is significant when compared to a supervised transformer-based method. This adds credence to the strength and reliability of our proposed framework and pre-training strategy.
\end{abstract}

\section{Introduction}
Chest X-rays (CXRs) are the preferred imaging modality for various clinical diagnostic tasks. This is supported by the fact that acquiring CXR is relatively cheap and the x-ray facilities are widely available. 
Since the pandemic, chest x-rays have been the baseline radiological investigation method for Coronavirus disease (covid-19) screening, particularly for symptomatic or covid-19 positive patients presented to the emergency department, as the Coronavirus disease is manifest in an acute respiratory illness affecting the lungs.  
One of the most widely used tests for covid-19 detection is the Reverse Transcription-Polymerase Chain Reaction (RT-PCR) test. However, the turnaround time for RT-PCR results is significant (up to 24 hours), whereas, CXR-based screening is faster and more helpful for observing disease progression. Initially, pediatric patients were thought to be immune to the virus. However, over multiple mutations, more than 7 million children have tested positive for covid-19 in the United States alone \cite{zachariah2022covid}. 
With an increasing burden on healthcare facilities during the pandemic, where people of all ages are affected by the coronavirus, CXR has proved very useful to triage covid-19 and pneumonia in a resource-constrained environment. The success of this diagnostic tool would be further boosted by an expert computer-aided diagnosis (CAD) system that can effectively process CXRs for pathology detection, thus managing to mitigate the growing workload for radiologists.   



Traditionally, medical diagnosis has focused on having a lower false negative rate (FNR) even at the cost of a higher false positive rate (FPR) \cite{woloshin2020false}. The pandemic, with immense stress on healthcare resources, required an additional focus on lowering the FPR. The recent success of machine learning, in particular deep learning algorithms, for automated image analysis has advanced the field of computer-aided diagnosis. The availability of convolutional neural network (CNN) architectures has fueled the development of innovative algorithms for optimizing various performance metrics like the FPR \cite{kim2021deep}. The other challenges of the development and clinical translation of deep learning methods are the scarcity of quality labels, limited data for rare diseases (particularly for pediatrics), and lack of generalization for handling imaging data drift from different clinical sites and devices. Foundation medical image analysis models, similar to those trained on large-scale domain corpora in natural language processing \cite{gu2021domain}, could help address the challenges for image-based CAD. Pre-training strategies for such models require massive amounts of training data, which is a challenge in the medical domain. For the protection of patient privacy, clinical data needs to be securely stored, limiting its availability, and access to large-scale clinical data sets.  However, within the clinical sites, unannotated data is increasing every day. Therefore, there is an urgent need to develop innovative methods that could learn from this data for better patient outcomes.             

Vision transformers (ViT) with large-scale training data achieved state-of-the-art performance in vision-related tasks like classification, detection, and semantic segmentation \cite{han2022survey,parvaiz2022vision}. The attention module in ViT can help identify the correct features for the downstream domain-specific tasks, such as pathology classification or image segmentation, compared to the standard CNN-based architectures. This aspect of transformers could help to optimize the parameters of interest, such as True Negative Rate (TNR) with reduced FPR. 

Group Masked Model Learning (GMML) is a pre-training strategy that can leverage sizeable unlabeled data sets to facilitate effective fine-tuning of the ViT \cite{atito2022gmml} for downstream tasks. 
Herein, we adopt the GMML strategy for domain-specific self-supervised pre-training using CXR data, 
thus ascertaining the feasibility of our proposed methodology for CXR-based diagnosis and image analysis in the clinical domain. Particularly, we show that the learned representation is successfully transferred to challenging small-scale unseen pediatric data with a significant performance in covid-19 detection.

\noindent \textbf{Our Contributions} are summarized as follows-
\begin{enumerate}
    \item We propose the use of GMML for self-supervised pre-training of ViT on chest x-rays (SS-CXR). 
    \item We further establish that our strategy helps a diverse set of domain-specific downstream tasks, including pathology classification and delineation of the boundaries between lung and background for segmentation.
    \item We conduct extensive fine-tuning experiments on CXRs on both seen and unseen data distributions (to assess the relevance of the proposed method in practical clinical scenarios) for multi-class classification, binary classification, and semantic segmentation.
    We establish the importance of pre-training transformers on the domain data for improved performance. 
    \item We are the first to show that a single model achieves a high degree of accuracy to classify healthy patients from pneumonia and covid-19 on one of the largest benchmark CXR data.
    \item To the best of our knowledge, we are the first to show that self-supervised pre-training on adult CXR helps learn useful representations for covid-19 detection in pediatric chest scans, traditionally an underrepresented class in most data sets.   
\end{enumerate}

\begin{figure*}[ht]
    \centering
    \includegraphics[width=0.80\linewidth]{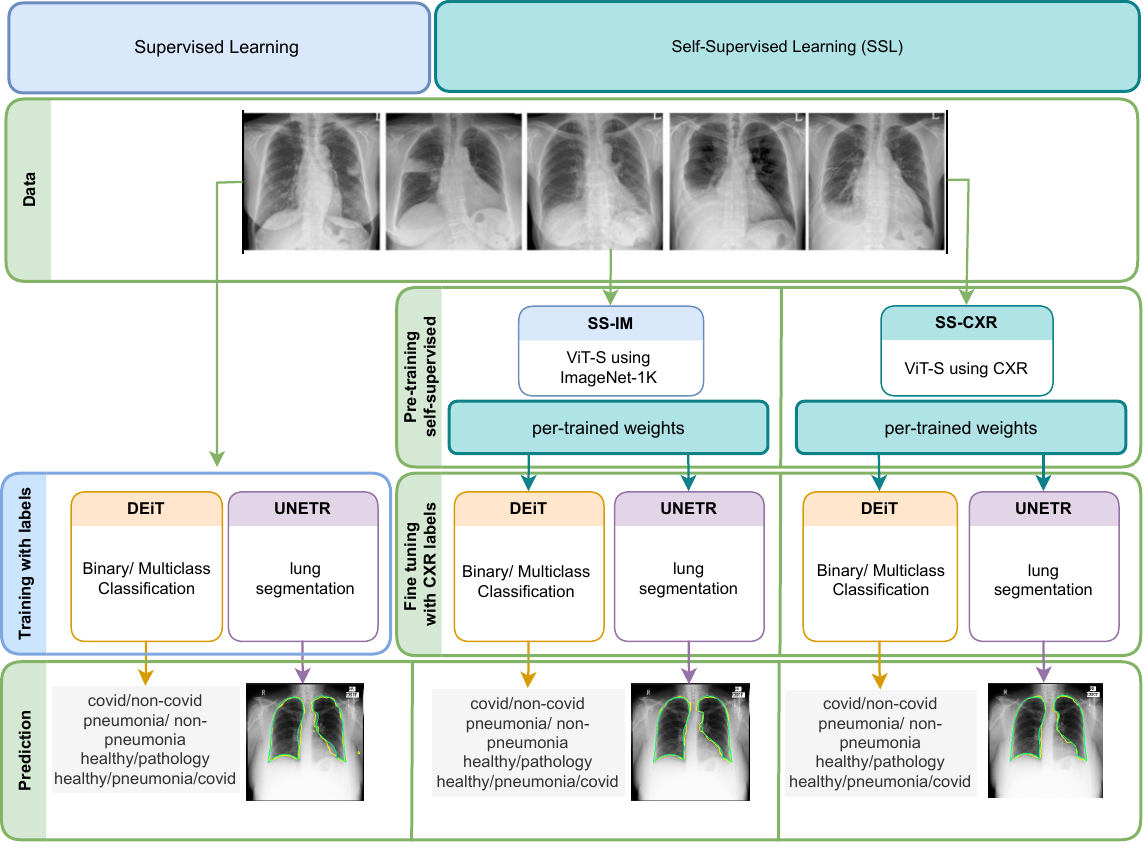}
    \caption{Proposed self-supervised ViT-based framework for chest x-ray images (SS-CXR). The architecture uses GMML-based self-supervised pre-training followed by fine-tuning of the state-of-the-art ViT-S frameworks:DEiT (for classification) and UNETR (for segmentation). We tackle multi-class classification and lung segmentation in the downstream tasks with significant performance.}
    \label{fig:GMMLtransformer}
\end{figure*}

\section{Related Works} 

The popularity of self-supervised learning (SSL) has led to several self-supervised pre-training methods over the years. Earlier works designed a range of pretext tasks, for example, reconstruction of the noisy or perturbed image via autoencoding~\cite{kramer1991nonlinear}, denoising autoencoder~\cite{vincent2008extracting}, predicting the order of the patches~\cite{noroozi2016unsupervised}, recovering color of grayscale image~\cite{zhang2016colorful}, predicting missing region via image inpainting~\cite{pathak2016context}, predicting rotation angle of image~\cite{gidaris2018unsupervised}, adversarial loss for masked patches~\cite{trinh2019selfie}. More recent methods focus on contrastive learning framework~\cite{oord2018representation,wu2018unsupervised,hjelm2018learning,bachman2019learning,
chen2020simple,NEURIPS2020_29539ed9,grill2020bootstrap}. The goal of the contrastive learning is to minimize the distance between two augmented views of the same image and maximize the distance between views from different images. 

Over the last years, due to the popularity of vision transformers~\cite{dosovitskiy2020image}, several SSL works start using the transformer's backbone. The basic form of masked image modeling for SSL pre-training has been explored in iGPT~\cite{chen2020generative} and ViT~\cite{dosovitskiy2020image}, with the recovery of quantized color space for each patch using GPT and BERT tasks. These SSL frameworks, both using CNNs and ViTs, have shown great potential for self-supervised pre-training, but none of them are able to outperform supervised pre-training, leading to ViT being trained on JFT-300M Google internally labeled dataset~\cite{dosovitskiy2020image}. Self-supervised vIsion Transformer (SiT)~\cite{atito2021sit} is a seminal work in the space of BERT-like masked image models via a simple masked autoencoder framework. 
Importantly, SiT demonstrated that it is possible to train data-hungry ViTs on small datasets from scratch. 
One of the limitations of SiT-based masked autoencoding frameworks is a relatively slow convergence. 


Self-supervised pre-training has been gaining appeal in the medical image domain. 
Inspired by the UNET architecture \cite{unet}, a transformer-based model was proposed for medical image analysis, particularly segmentation \cite{unetr}. The model used a self-supervised pre-training strategy and was evaluated on the body organ segmentation challenge datasets with state-of-the-art performance. A self-supervised contrastive learning strategy was used for covid-19 prognosis \cite{sriram2021covid}. The models were able to outperform radiologists in predicting mortality for covid-19 patients. In another study, a self-supervised pre-training strategy was applied to magnetic resonance imaging for knee abnormality classification \cite{atito2022sb}. It was shown that if appropriate strategies are used, such methods can be utilized in limited data regimes. Herein, we take a leap towards developing a model along the lines of a foundation model for CXR data. Our proposed strategy is compared with standard supervised learning as well as ImageNet-based pre-training to demonstrate the merits and feasibility of the proposed approach and its potential for achieving better patient outcomes.


\section{Self-Supervised Pre-training for Medical Imaging}

Self-Supervised learning is an alternative to supervised learning. It operates upon the input data without output (ground truth) labels. The fundamental idea is to generate supervisory signals from the unstructured data by defining a pretext task, where a machine learning framework learns the underline structure of the data to solve the task. To that end, SSL methods can be trained using much bigger data that is less affected by human labeling bias and noise within the data. 
In this work, we introduce SS-CXR, a general self-supervised vision transformer framework for chest x-rays. Our proposed SS-CXR framework uses vision transformers \cite{pmlr-v139-touvron21a} and GMML, explained in Section \ref{sec:GMML}. The learned representation from the pre-training task is used for various domain-specific downstream tasks (Section \ref{sec:downstream}).  Figure~\ref{fig:GMMLtransformer} shows the system diagram of our proposed approach. 

\subsection{Group Masked Model Learning (GMML)}
\label{sec:GMML}

Masked image modeling, which in principle is similar to the Masked Language Modeling (MLM) used in BERT~\cite{devlin2018bert}, was first proposed in SiT \cite{atito2021sit} and employed in several recent vision \cite{bao2021beit,atito2021mc,xie2022simmim} and medical \cite{atito2022sb,chen2022masked} works. 
The main idea of GMML is to corrupt a group of connected patches representing a ``significant'' part of a given visual input and recover them by learning a model. The underlying hypothesis is that, by recovering the corrupted parts of a given visual input from the uncorrupted parts based on the context of the whole visual field, the network will implicitly learn the notion of visual integrity. Intuitively the network is able to recover the missing information only if it learns the characteristic properties of visual stimuli corresponding to specific actions impacting the visual input. The weights of the learned model can then be employed as an initialization point for a related (domain-specific) downstream task. 

\begin{figure}[ht]
    \centering
    \includegraphics[width=\linewidth]{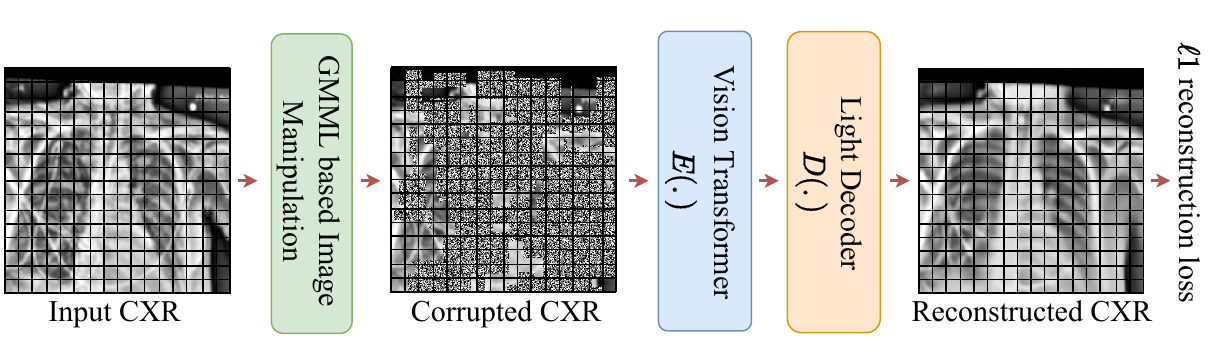}
    \caption{GMML strategy for SS-CXR showing x-ray image manipulation and vision transformer-based encoder trained with reconstruction loss for reconstructing the CXR image.}
    \label{fig:GMMLtransformerarch}
\end{figure}

There are several ways to corrupt a group of connected patches. For example, replacing the randomly selected patches with zeros, noise, or with patches from other medical images. Note that groups of corrupted patches are selected randomly, i.e., the patches with vital organs and the patches with the background are manipulated with equal probability and the network is required to learn to reconstruct the information corresponding to the whole medical image with equal importance. This equips the network for generalizing well to unseen data, as demonstrated by the results for pediatric cases (Section \ref{finetune_drift}).
For CXR reconstruction, we propose to use the ViT as a group masked auto-encoder, i.e., visual transformer auto-encoder with GMML. By analogy to auto-encoders, our network is trained to reconstruct the input image through the output tokens of the transformer. The schematic for pre-training using CXR data is shown in Figure \ref{fig:GMMLtransformerarch}. 

Vision transformer receives as input a sequence of patches obtained by tokenizing the input x-ray image $\mathbf{x} \in \mathbb{R}^{H \times W}$ into $n$ flattened $2D$ patches of size $p \times p$ pixels, where $H$ and $W$ are the height and width resolution of the input x-ray and $n$ is the total number of patches. The $n$ patches are then flattened and fed to a linear projection, which is then passed to the ViT. 

To pre-train transformers as group masked auto-encoder, the GMML manipulated x-ray image $\mathbf{\hat{x}}$ is first obtained by randomly replacing a significant part of the image with noise (70\% of the input CXR image) or with patches from another CXR image (35\%). 

The reconstructed x-ray image $\mathbf{\bar{x}}$ is obtained by passing the corrupted image $\mathbf{\hat{x}}$ to the transformer encoder $E(.)$ and feeding the output to a light decoder $D(.)$. The decoder consists of 3 fully connected layers; the first two with $2048$ neurons and GeLU non-linearity each, and the last bottleneck layer with $256$ neurons, followed by a transposed convolution to return back to the medical image space. After the pre-training stage, the light decoder $D(.)$ is dropped and only the encoder $E(.)$ is used for the downstream task. The pseudo-code for SS-CXR is presented in Algorithm 1. The importance of using the representation before the nonlinear projection i.e., decoder head, is due to the loss of information induced by the reconstruction loss. Particularly, the decoder head generally learns task-specific features i.e., CXR image reconstruction, which might rescind information that may be useful for the downstream task.

For the x-ray reconstruction task, we use the $\ell1$-loss between the original and the reconstructed image as shown in Equation \ref{eq:l1-pixel}. We compute the loss only on the corrupted pixels, similar to \cite{devlin2018bert,atito2022gmml}.

\begin{align}
\label{eq:l1-pixel}
\mathcal{L}(\mathbf{W})_{\rm recons} &= \sum_k^N \left( \sum_i^H \sum_j^W \mathbf{M}_{i,j}^k \times | \mathbf{x}_{i,j}^k - \mathbf{\bar{x}}^k_{i,j} | \right),
\\
\mathbf{M}_{i,j} &= 
\begin{cases}
    1,              & \text{if } \mathbf{x}_{i,j} \text{ is manipulated}\\
    0,              & \text{otherwise,}
\end{cases}
\end{align}
where $\mathbf{W}$ denotes the parameters to be learned during training, $N$ is the batch size, and $\mathbf{M}$ is a binary mask with 1 indicating the manipulated pixels.

\begin{algorithm}[!th]
\caption{\mbox{SS-CXR}: PyTorch-style pseudo-code.}
\label{algo:sitml}
\begin{lstlisting}
#E(.): Encoder network, i.e. vision transformer
#D(.): light decoder 

for X in dataloader: # load batch with N samples
    X_clean = augment(X) # N x H x W
    X_corrupt = corrupt(X_clean) # N x H x W
    
    # compute projections
    x_recons = D(E(X_corrupt)) # N x H x W
    
    # compute reconstruction loss
    loss = L1Loss(x_recons, X_clean)    
    
    # optimization step
    loss.backward()
    optimizer.step()


\end{lstlisting}
\end{algorithm}

\subsection{Classification and Segmentation Downstream Tasks}
\label{sec:downstream}
 
 After the self-supervised pre-training step, the lightweight decoder is removed from the SS-CXR framework and replaced with task-specific layers for the downstream task. We hypothesize that the pre-training from unlabeled CXRs using our proposed strategy (SS-CXR) learns substantial knowledge representation for task-specific fine-tuning of network weights. The fine-tuning for the downstream tasks is performed in a supervised manner using state-of-the-art classification (DEiT \cite{pmlr-v139-touvron21a}) and segmentation models (UNETR \cite{unetr}). We demonstrate the effectiveness of pre-training stage with two downstream tasks, i.e., lung disease classification and lung segmentation using the CXR data. 

\noindent \textbf{Classification Task:}
For the classification task, a trainable vector (i.e., class token) is appended to the input sequence of the patch tokens and passed through the transformer encoder. A classification head is added to the output of the transformer encoder corresponding to the class token.  A linear layer that projects the features into the number of classes serves as the classification head.  
We perform several downstream classification tasks including binary and multi-class classifications. For the binary classification, we fine-tune the pre-trained models to classify CXR into healthy/pathology, covid/non-covid, and pneumonia/non-pneumonia. For healthy/pathology classification, pathological cases represented both covid-19 and pneumonia. For the multi-class classification, we fine-tune the pre-trained models to classify CXR into healthy, pneumonia, and covid-19. Further discussion and implementation details are in Section \ref{sec:exp_classification}.
\noindent \textbf{Segmentation Task:}
Drawing inspiration from the recent success of U-Net architecture in medical image segmentation \cite{unet}, a U-shaped architecture called UNETR was proposed employing transformers \cite{unetr}. In that work, ViT was used in the encoder to capture the global multi-scale information, and transposed convolutional layer as a decoder to generate the segmentation map. Herein, we modify this framework and fine-tune ViT-S in the UNETR encoder initialized with weights from the pre-training using the SS-CXR training paradigm. 
Refer to Section \ref{sec:lungsegmentation} for further discussion and implementation details.

\section{Experimental Results}
\subsection{Datasets}
We aim to show that the domain-specific representation learned during the self-supervised pre-training stage using our proposed strategy (SS-CXR) has the potential to solve frequent clinical tasks (classification and segmentation). To this end, we use one of the biggest collections of chest x-rays with covid-19 and pneumonia patients, the COVIDxCXR-3 benchmark dataset\footnote{\url{https://www.kaggle.com/datasets/andyczhao/covidx-cxr2}} \cite{Covidnet, pavlova2022covidx}. The data are diverse and collected from multiple countries and clinical sites including both adult and some pediatric patients. Hence, is well-suited to show the robustness of our self-supervised pre-training strategy.   
This data is a collection of multiple datasets including BIMCV \cite{bimcv}, STONYBROOK \cite{stony}, RSNA pneumonia detection \cite{rsna}, RSNA-RICORD \cite{ricord}, SIRM \cite{SIRM}, FIG-1\cite{fig1}, ACTMED \cite{actmed} and COHEN\cite{cohen}. 
We have 31,493 chest x-rays 
from patients whose ages ranged from 0 to 100 years. The number of pediatric patients is significantly small, 0.2\% of patients were $<$ 18 years \cite{pavlova2022covidx}. The data were split into a train (31,093 x-rays) and test (400 x-rays) sets . In the training data set, 8596 samples are normal, 5503 correspond to pneumonia, and 15994 are labeled as Covid-19. The self-supervised pre-training stage used all the data without any ground-truth label.  For our experiments, COHEN and ACTMED data were not used for pre-training and fine-tuning and were exclusively used as a test set to evaluate shifts in data distribution (arising from a different site/scanner). Similarly, the test data (200 x-rays from RICORD and RSNA each) were excluded from pre-training and fine-tuning. Details of the data are presented in 
in the supplementary Section \ref{section:dataset-clf} 

Further, to evaluate the robustness of the representation learning, we curated a dataset for pediatric CXR in-house to test our proposed paradigm. After approval from the Internal Review Board (IRB), 189 posterior-anterior CXRs were acquired from our medical center's Picture Archiving and Communication System (PACS). The data comprised x-rays from 90 healthy subjects and 99 patients diagnosed with covid-19. The age range of the healthy subjects is 0 to 21 years, where most patients were between 0-1 year old. For covid-19 patients, age varies between 0 to 23 years, where most patients were 0-2 years old (more details in supplementary Section \ref{section:cnhdataset}).


For lung segmentation task in chest x-rays, we used the Shenzhen dataset \cite{jaeger2013automatic}, Montgomery County dataset \cite{candemir2013lung} and Belarus Tuberculosis Portal \footnote{\url{http://tuberculosis.by}}. Hence, we had a total of 801 CXRs, out of which we used 138 CXRs from the Montogomery site for testing.  
The numbers and splits used for the pre-training, fine-tuning, and testing stages of the experiments are presented in 
supplementary Section \ref{section:dataset-seg}.

\begin{table*}[t]
\centering
\caption{{Covid vs Non-Covid binary classification performance employing ViT-S. The RSNA \& RICORD training data (training data distribution) were used for fine-tuning, whereas all of COHEN, ACTMED, and pediatric data were only used for testing.}}
\label{tab:covid-non-covid}
\resizebox{1\linewidth}{!}{
\begin{tabular}{l|cccccc|cccccc|cccccc}
\hline
& \multicolumn{6}{c|}{\textbf{RSNA \& RICORD}}                          & \multicolumn{6}{c|}{\textbf{COHEN \& ACTMED}}                           & \multicolumn{6}{c}{\textbf{Pediatric Data}}  \\ \hline
\textbf{Testing Paradigm}
& \multicolumn{6}{c|}{Training Data Distribution}
& \multicolumn{6}{c|}{Unseen Data Distribution} 
& \multicolumn{6}{c}{Shifted Data Distribution} \\ \hline
\textbf{Non-Covid / Covid}
& \multicolumn{6}{c|}{200 / 200}
& \multicolumn{6}{c|}{404 / 295} 
& \multicolumn{6}{c}{90 / 99} \\ \hline
& TPR & FPR & FNR & 
\begin{tabular}[c]{@{}c@{}}AUC\\ PR\end{tabular} & 
\begin{tabular}[c]{@{}c@{}}AUC\\ ROC\end{tabular} & ACC 
& TPR & FPR & FNR & 
\begin{tabular}[c]{@{}c@{}}AUC\\ PR\end{tabular} & 
\begin{tabular}[c]{@{}c@{}}AUC\\ ROC\end{tabular} & ACC 
& TPR & FPR & FNR & 
\begin{tabular}[c]{@{}c@{}}AUC\\ PR\end{tabular} & 
\begin{tabular}[c]{@{}c@{}}AUC\\ ROC\end{tabular} & ACC \\ \hline
{ViT-S} \cite{Vit} & 0.975 & 0.025 & 0.190 & 0.972 & 0.977 & 89.25 & 0.740 & 0.260 & 0.319 &0.693 & 0.755 & 71.53 & 0.522 & 0.478 & 0.118 &0.619 & 0.702 & 68.42\\
{ViT-S + SS-IN} &\textbf{1}& \textbf{0} & \textbf{0.035} & 0.998 &0.998 &\textbf{98.25} & 0.762 & 0.237 & 0.389 & 0.650 & 0.723 & 69.81 & 0.855 & 0.144 &0.646 &0.642 &0.605 &59.26\\
\textbf{ViT-S + SS-CXR} &\textbf{1}& \textbf{0} & \textbf{0.035} & \textbf{0.999} & \textbf{0.999} & \textbf{98.25} & \textbf{0.779} &\textbf{0.220}&\textbf{0.308}&\textbf{0.726}&\textbf{0.789}& \textbf{74.24}& \textbf{0.952} & \textbf{0.048}& \textbf{0.059} & \textbf{0.997}& \textbf{0.997}& \textbf{94.73}\\ \hline
\end{tabular}
}
\end{table*}

\begin{table*}[t]
\centering
\caption{{Healthy vs Pathological binary classification employing ViT-S. The RSNA \& RICORD training data (training data distribution) were used for fine-tuning, whereas all COHEN, ACTMED, and pediatric data were only used for testing.}}
\label{tab:healthy-pathology}
\resizebox{1\linewidth}{!}{
\begin{tabular}{l|cccccc|cccccc|cccccc}
\hline
& \multicolumn{6}{c|}{\textbf{RSNA \& RICORD}}                          & \multicolumn{6}{c|}{\textbf{COHEN \& ACTMED}}                           & \multicolumn{6}{c}{\textbf{Pediatric Data}}  \\ \hline
\textbf{Testing Paradigm}
& \multicolumn{6}{c|}{Training Data Distribution}
& \multicolumn{6}{c|}{Unseen Data Distribution} 
& \multicolumn{6}{c}{Shifted Data Distribution} \\ \hline
\textbf{Healthy / Pathology}
& \multicolumn{6}{c|}{100 / 300}
& \multicolumn{6}{c|}{404 / 295} 
& \multicolumn{6}{c}{90 / 99} \\ \hline
& TPR & FPR & FNR & 
\begin{tabular}[c]{@{}c@{}}AUC\\ PR\end{tabular} & 
\begin{tabular}[c]{@{}c@{}}AUC\\ ROC\end{tabular} & ACC 
& TPR & FPR & FNR & 
\begin{tabular}[c]{@{}c@{}}AUC\\ PR\end{tabular} & 
\begin{tabular}[c]{@{}c@{}}AUC\\ ROC\end{tabular} & ACC 
& TPR & FPR & FNR & 
\begin{tabular}[c]{@{}c@{}}AUC\\ PR\end{tabular} & 
\begin{tabular}[c]{@{}c@{}}AUC\\ ROC\end{tabular} & ACC \\ \hline
{ViT-S} \cite{Vit} &\textbf{0.980} & \textbf{0.020}&0.067 &0.995 & 0.985 & 94.50 & 0.438 & 0.562 &0.176 &0.559 &0.539 &\textbf{61.51}& 0.524 & 0.476 & 0.118& 0.620 &0.689 &68.42\\
{ViT-S + SS-IN} & \textbf{0.980} & \textbf{0.020} & 0.043 & 0.996 & 0.986 & 96.25 & 0.332 & 0.668 & \textbf{0.078}& 0.551 &0.672 &58.08 &0.711&0.299 &0.424 &0.693 &0.661 &64.02\\
\textbf{ViT-S + SS-CXR} &0.970 &0.030&\textbf{0.030} &\textbf{0.997}& \textbf{0.992} &\textbf{97.00} & \textbf{0.463} & \textbf{0.537}& 0.153 & \textbf{0.665} & \textbf{0.679} &61.08   & \textbf{0.952} &\textbf{0.048} &\textbf{0}& \textbf{1}&\textbf{1}&\textbf{97.36}\\ \hline
\end{tabular}
}
\end{table*}

\begin{table*}[!ht]
\centering
\caption{{Pneumonia/Covid and Pneumonia/Non-Pneumonia binary classification employing ViT-S variant of vision transformers}.}
\label{tab:multi-table}
\resizebox{0.8\linewidth}{!}{
\begin{tabular}{l|cccccc|cccccc}
\hline
& \multicolumn{6}{c|}{\textbf{RSNA \& RICORD} (Pneumonia vs Covid-19)}                          & \multicolumn{6}{c}{\textbf{RSNA \& RICORD} (Pneumonia vs Non-Pneumonia)}                         \\ \hline
\textbf{Testing Paradigm}
& \multicolumn{12}{c}{Training Data Distribution} \\ \hline

& \multicolumn{6}{c|}{100 / 200}
& \multicolumn{6}{c}{300 / 100} 
 \\ \hline
& TPR & FPR & FNR & 
\begin{tabular}[c]{@{}c@{}}AUC\\ PR\end{tabular} & 
\begin{tabular}[c]{@{}c@{}}AUC\\ ROC\end{tabular} & ACC 
& TPR & FPR & FNR & 
\begin{tabular}[c]{@{}c@{}}AUC\\ PR\end{tabular} & 
\begin{tabular}[c]{@{}c@{}}AUC\\ ROC\end{tabular} & ACC \\ \hline
ViT-S \cite{Vit} & 0.960 & 0.040 & 0.040 & 0.994 &0.989 & 96 & 0.973 & 0.027 & 0.080 &0.961 &0.978 & 96.00\\

ViT-S + SS-IN & \textbf{1}& \textbf{0} & 0.025 & 0.999 & 0.999 & 98.30 & \textbf{0.990} & \textbf{0.010} &0.060 & 0.984 & 0.989 & \textbf{98.00}\\

\textbf{ViT-S + SS-CXR} & \textbf{1}& \textbf{0} & \textbf{0.005} & \textbf{1} & \textbf{0.999} & \textbf{99.67} & \textbf{0.990} & \textbf{0.010} &\textbf{0.050} & \textbf{0.988} & \textbf{0.995} & \textbf{98.00}\\ \hline
\end{tabular}
}
\end{table*}

\begin{table*}[t]
\centering
\caption{{Multiclass (Healthy vs Pneumonia vs Covid) classification employing ViT-S variant of vision transformers. Class labels are represented as 0: healthy, 1: pneumonia, 2: covid-19.}}
\label{tab:multi-class}
\resizebox{0.97\linewidth}{!}{
\begin{tabular}{l|ccccc|ccccc|ccccc|c}
\hline
& \multicolumn{16}{c}{\textbf{RSNA \& RICORD}} \\ \hline
\textbf{Testing Paradigm}
& \multicolumn{16}{c}{Training Data Distribution} \\ \hline
\textbf{Healthy (0)/ Pneumonia (1)/ Covid-19 (2)}
& \multicolumn{16}{c}{100 / 100 / 200}
\\ \hline

& TPR\_0 & FPR\_0 & FNR\_0 & 
\begin{tabular}[c]{@{}c@{}}AUC\\ PR\_0\end{tabular} & 
\begin{tabular}[c]{@{}c@{}}AUC\\ ROC\_0\end{tabular}& TPR\_1 & FPR\_1 & FNR\_1 & 
\begin{tabular}[c]{@{}c@{}}AUC\\ PR\_1\end{tabular} & 
\begin{tabular}[c]{@{}c@{}}AUC\\ ROC\_1\end{tabular}  & TPR\_2 & FPR\_2 & FNR\_2 & 
\begin{tabular}[c]{@{}c@{}}AUC\\ PR\_2\end{tabular} & 
\begin{tabular}[c]{@{}c@{}}AUC\\ ROC\_2\end{tabular} & ACC\\ \hline
ViT-S \cite{Vit} & \textbf{0.990} & \textbf{0.010}& 0.077& 0.953 &0.987 &0.890 &0.110 &0.013& 0.958 &0.978 &0.910 &0.090 &0.015 &0.984 &0.987& 92.50\\

ViT-S + SS-IN & \textbf{0.990} & \textbf{0.010}& 0.047 &965 &0.990 &0.940 &0.060 & \textbf{0.007}& 0.986 &0.993 & \textbf{0.955} & \textbf{0.045} & \textbf{0} & 0.997 &0.995 &\textbf{96.00}\\

\textbf{ViT-S + SS-CXR} & 0.980 & 0.020 &\textbf{0.043}& \textbf{0.972} & \textbf{0.991} & \textbf{0.950} & \textbf{0.050}& \textbf{0.007} &\textbf{0.987} & \textbf{0.995} & \textbf{0.955} & \textbf{0.005} & 0.005 & \textbf{0.997}  & \textbf{0.997}& \textbf{96.00} \\ \hline
\end{tabular}
}
\end{table*}

\subsection{Pathology Classification}
\label{sec:exp_classification}
 Binary and multi-class classification tasks were performed to identify covid-19, pneumonia, and healthy subjects. The selection of these tasks was dependent upon the availability of appropriate data. For CXR classification, model fine-tuning was performed on an image size of 256$\times$256 with AdamW optimizer \cite{adamw}, a weight decay of $0.05$, and a learning rate of ${5e^{-4}}$ for 100 epochs. For our experiments we used the ViT-S (with an embedding dimension of 384, and 22M parameters) and our training strategies include: 1) fully supervised ViT-S, 2) GMML-based self-supervised pre-training on ImageNet-1K followed by fine-tuning (SS-IN), and 3) GMML-based self-supervised pre-training on CXRs followed by fine-tuning (SS-CXR). For ImageNet-1K pre-training, we used the ImageNet subset with 1000 object classes and 1,281,167 training images.  
 The classification performance is evaluated using the True Positive Rate (TPR), False Positive Rate (FPR), False Negative Rate (FNR), Area under the Precision-Recall curve (AUC-PR), Area under the Receiver Operator Curve (AUC-ROC), and accuracy (ACC) \cite{biship2007pattern}. 

\noindent \textbf{Binary Classification:}
We experiment with the classification of covid vs non-covid (Table \ref{tab:covid-non-covid}), pathology (covid and pneumonia) vs healthy  (Table \ref{tab:healthy-pathology}), and pneumonia vs healthy (Table \ref{tab:multi-table}). We simulated various testing paradigms and reported results on three different test data. The CXR-3 test data has samples only from RSNA \& RICORD sites (Supplementary Section \ref{section:dataset-clf}) which are also part of the training data. The training data has been used in the fine-tuning step, hence when testing on RSNA \& RICORD data, we refer to the testing paradigm as training data (owing to similar sites for train and testing) distribution. We exclusively held the COHEN \& ACTMED data from pre-training and fine-tuning and hence, were used to evaluate the data distribution shift (owing to different sites and scanners) in the unseen data distribution paradigm. Similarly, none of the in-house pediatric data were used for pre-training or fine-tuning and comprised the shifted data distribution (owing to the difference in the chest anatomy of adults and children) testing paradigm. Our results show a consistent improvement in quantitative evaluation metrics for our proposed SS-CXR approach. 
This phenomenon is consistent across various data subsets, and data drifts presented in sections \ref{finetune_drift} and \ref{testtime_drift}, showing the strength of our proposed framework.

\noindent \textbf{Multi-class Classification:}
The effectiveness of the representation learning using SS-CXR is further evaluated by performing a multi-class classification task. CXRs are classified as healthy, covid-19, and pneumonia. The results are presented in Table \ref{tab:multi-class}. 
Our trained model, ViT + SS-CXR, achieved an accuracy of 96\% which is close to the classification accuracy achieved for binary classification (pneumonia/non-pneumonia: 98\%, covid/non-covid: 98.25\%, and healthy/pathology: 97\% ). It is an indication that using the multi-class classification approach it is possible to achieve almost a hierarchical binary classification performance with a single model, thus making faster and more accurate inferences with fewer learnable parameters.

\subsubsection{Data Distribution Shift during Fine Tuning}\label{finetune_drift}

The CXR is a 2D image of the chest region for adults and children, but the anatomy varies significantly for both categories. This difference is more significant when we consider the case of neo-natal CXR, where several organs and bone structures have different anatomy. Further, it is difficult to keep young children stationary when x-rays are captured in multiple fields of view and orientations \cite{mansoor}. The pediatric data in general and for covid-19 are rare to find publicly and are generally smaller compared to data available for older subjects. This has traditionally limited the application of deep learning methods to pediatric data and rare diseases in pediatric populations. With the proposed SS-CXR training strategy, the aim is to develop a self-supervised method suited for these smaller data regimes, expanding the application of deep learning to more challenging clinical applications. The classification performance on pediatric in-house CXR data helps to add credence to the superior performance of the ViT-S + SS-CXR model on small datasets (in the order of magnitude of a couple of hundred). The  classification accuracy achieved (94.73\% for covid/non-covid detection in Table \ref{tab:covid-non-covid}) is significantly higher than using a fully supervised ViT-S model (an increase of $\sim$26\%) and SS-IN training (an increase of $\sim$35\%). It should be noted that none of this data was used during pre-training or fine-tuning. We argue that the representation learned using the SS-CXR strategy aided in a successful application of the ViT-based deep learning model for such a small dataset.

\subsubsection{Unseen Data Distribution at Test Time}\label{testtime_drift}

The data from COHEN \cite{cohen} and ACTMED \cite{actmed} sites were excluded from the pre-training and fine-tuning steps. This is to simulate the scenario frequently encountered in the medical imaging domain, namely the need to use the designed system on data collected in a different environment. This often results in a drift in the intensity distribution in scans due to differences in protocols observed at medical centers. Hence, machine learning models performing well when trained and evaluated on data from a particular clinical center or scanner, fail when tested on data from a different clinical center or scanner.  

For binary classification (covid/non-covid: Table \ref{tab:covid-non-covid} 
and healthy/pathology: Table \ref{tab:healthy-pathology}) a significant gap in classification performance is observed for the training data (ACC: 98.25\%, FPR: 0) and unseen data distribution paradigms (ACC: 74.24\%, FPR: 0.220). 
However, ViT-S + SS-CXR performance is significantly better in some cases by almost a factor of 2 (For instance in Table \ref{tab:covid-non-covid}, FNR for covid/non-covid detection in pediatric data: 0.059 whereas FNR for ViT-S: 0.118, similarly ACC for healthy/pathology detection in Table \ref{tab:healthy-pathology} for pediatric data: 97.36\%, whereas ACC for ViT-S: 68.42\%). This phenomenon can be explained by the pre-training strategy, where the idea was to learn the generic topology of CXR and the model was able to perform better than a fully supervised transformer trained from scratch even for unseen and shifted data distributions. 

\begin{table}[ht]
\centering
\caption{{Performance analysis for semantic lung segmentation with fully supervised and self-supervised variants. IoU: Intersection over union, HD: Hausdroff distance.}}
\label{tab:segmentation_results}
\resizebox{0.99\linewidth}{!}{
\begin{tabular}{l|ccc}
\hline
& \multicolumn{3}{c}{MONTGOMERY (138 scans)} \\ \hline

& \begin{tabular}[c]{@{}c@{}}\textbf{IoU}\\ (Jaccard index)\end{tabular}& 
\begin{tabular}[c]{@{}c@{}}\textbf{DICE}\\ (F1 Score)\end{tabular} & 
\begin{tabular}[c]{@{}c@{}}\textbf{Surface Distance}\\ (HD-95)\end{tabular}\\ \hline
UNETR \cite{unetr} & 0.8566 & 0.9227 & 16.70\\

UNETR+ SS-IN & 0.8963 & 0.9453 & \textbf{9.07}\\

\textbf{UNETR + SS-CXR} & \textbf{0.9160} & \textbf{0.9561} & 9.73 \\ \hline
\end{tabular}
}
\end{table}

\begin{figure*}[ht]
\begin{center}  
\includegraphics[width=0.95\textwidth]{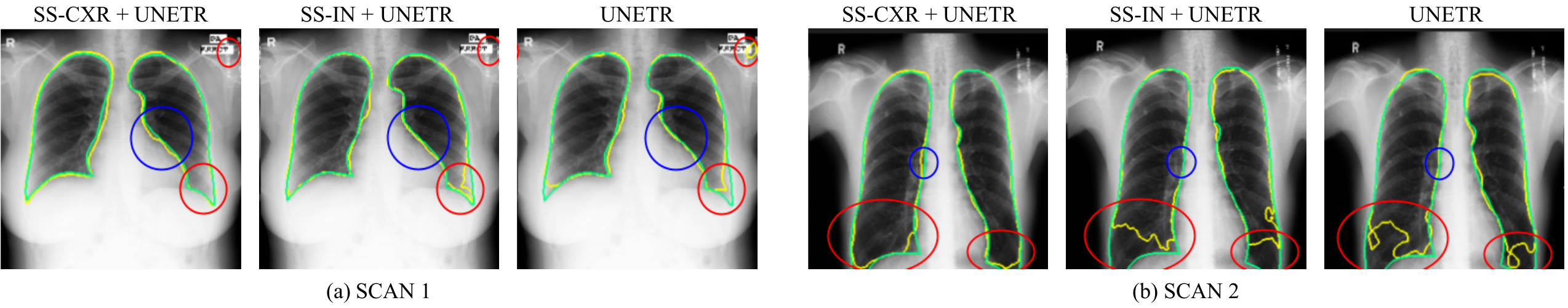}
\end{center}
\caption{\textbf{Visual analysis of lung segmentation}. The yellow polygons represent the predicted segmentation for two scans from the test set (Montgomery dataset \cite{candemir2013lung}). The green polygon traces the outline of the ground truth. The red and blue circles highlight the areas where there is a significant difference in predictions from different models. A zoomed version is available in 
supplementary Section \ref{section:appendix-seg}. }
\label{fig:segmentation}
\end{figure*}
\begin{figure}[ht]
\begin{center}  
\includegraphics[width=0.75\linewidth]{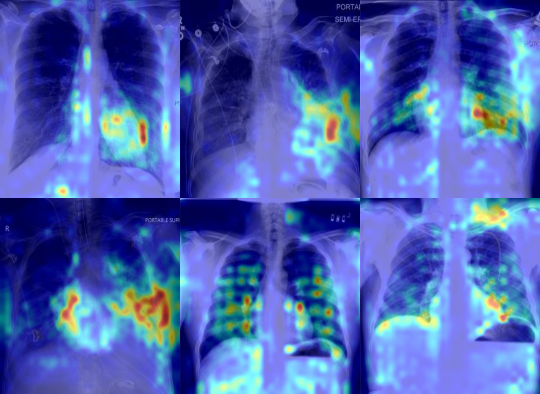}
\end{center}
\caption{{Heat maps representing regions in CXR the transformer network is attending the most for model interpretation and qualitative assessment.}}
\label{fig:heatmaps}
\end{figure}

\begin{table*}[t]
\centering
\caption{{Performance comparison for covid/non-covid classification in terms of accuracy (ACC) and true positive rate (TPR) of SS-CXR with human baseline, supervised ViT, and supervised CNNs with and without pretraining.}}
\label{tab:performance}
\resizebox{1\linewidth}{!}{
\begin{tabular}{l|c|ccccccccccc}
\hline
& \multicolumn{1}{c|}{\textbf{Human Baseline}} & \multicolumn{11}{c}{\textbf{Deep Learning Algorithms on CXR Test Data}} \\ \hline
 &  & \multicolumn{1}{c|}{\textbf{SSL}} & \multicolumn{6}{c|}{\textbf{Supervised (without pretraining)}} & \multicolumn{4}{c}{\textbf{Supervised (with pretraining)}}\\ \hline
 
&\begin{tabular}[c]{@{}c@{}}Thoracic\\ Radiologist \cite{hwang2021covid}\end{tabular}& SS-CXR & 
ViT-S & \begin{tabular}[c]{@{}c@{}}DenseNet-169\\ \cite{comppaper}\end{tabular} & \begin{tabular}[c]{@{}c@{}}EfficientNet-B2\\ \cite{comppaper}\end{tabular} & \begin{tabular}[c]{@{}c@{}}InceptionResnet-V2\\ \cite{comppaper}\end{tabular}& 
\begin{tabular}[c]{@{}c@{}}Inception-V3\\ \cite{comppaper}\end{tabular}& \begin{tabular}[c]{@{}c@{}}VGG-16\\ \cite{chodunsky_2021} \end{tabular}&\begin{tabular}[c]{@{}c@{}}DenseNet-121\\ (CXR) \cite{chodunsky_2021}\end{tabular}  & \begin{tabular}[c]{@{}c@{}}ResNet50-V2\\ (Bit-M) \cite{chodunsky_2021}\end{tabular} & \begin{tabular}[c]{@{}c@{}}Covid-Net\\ CXR-2 \cite{pavlova2022covid}\end{tabular}  & \begin{tabular}[c]{@{}c@{}}VGG-19\\ (ImageNet) \cite{chodunsky_2021}\end{tabular}\\ \hline
\textbf{ACC} & 64.37\footnotemark &\textbf{98.25} & 89.25 &98.15 & 97.60 & 97.55 & 97.50& 97.50 & 96.50 & 96.50 & 96.30 & 96.25\\

\textbf{TPR} & 0.645 & \textbf{1} & 0.975 &0.970 & 0.960 & 0.959 &0.952 & 0.950 & 0.935& 0.930 &0.955 &0.925 \\ \hline
\end{tabular}
}
\end{table*}

\subsection{Lung Segmentation}
\label{sec:lungsegmentation}
Semantic segmentation is the process of classifying pixels of an image into given categories. We focused on lung segmentation, i.e., the model classifies each pixel in the x-ray image into the foreground (lung) or background. Mean-IOU (intersection over union), Sorensen dice coefficient (F1-score), and surface distance (Hausdorff Distance) are the metrics used for the evaluation of lung segmentation. We fine-tune the self-supervised pre-trained ViT-S\cite{dosovitskiy2020image} within the UNETR framework on the Shenzhen \cite{jaeger2013automatic} and Belarus datasets and report results using the above metrics on the Montgomery (test) dataset \cite{candemir2013lung}. In this experiment, the ViT-S weights used are the same as those employed for the classification task and are obtained by the self-supervised pre-training using the CXR data without labels. For fine-tuning, we use an image size of 256 $\times$ 256 and trained on 663 samples from the Shenzhen \cite{jaeger2013automatic} and Belarus data. The AdamW optimizer\cite{adamw} with a weight decay $1e^{-5}$ and a learning rate ${5e^{-4}}$ was used for 400 epochs. 

The segmentation performance is presented 
in Table \ref{tab:segmentation_results}, where self-supervised training on CXR outperforms the fully supervised UNETR. A significant improvement is evident in the surface distance between the ground truth and the prediction (pre-trained models improve the surface distance by $\sim$46\%). The SS-CXR and SS-IN models focus on the boundaries due to learning better generic features seen regularly in the concept of chest X-rays. Figure \ref{fig:segmentation} shows that improvement in the surface distance helps in better segmentation outcomes.



\section{Discussion}

Table \ref{tab:performance} compares the SS-CXR framework with self-supervised (ViT-S) and fully supervised methods trained both with and without pre-training. We observe an improvement in performance in terms of both accuracy and TPR. Our proposed strategy performs significantly better than a thoracic radiologist in identifying covid-19 from CXRs, hence can be an efficient computer-aided diagnostic system. Apart from covid-19 classification, our method performs well to keep both the TPR and the TNR high, resulting in a better AUC-ROC score (Table \ref{tab:covid-non-covid}). This would lead to the real-world impact of lowering the hospitalization rates for patients  falsely classified as suffering from covid-19. Further, our framework differentiates bacterial pneumonia from viral covid-19 with a high accuracy of 99.67\% (Table \ref{tab:multi-table}).

For pediatric data without pre-training, the accuracy for the binary classification is low at 68.42\% (Table \ref{tab:covid-non-covid}). This can be attributed to the size of the dataset, which is very small and the model starts to overfit onto the train set and does not generalize well for the test set. The SS-CXR pre-training strategy on adult CXRs helps boost the classification accuracy of the pediatric data to an impressive rate of 94.73\% (Table \ref{tab:covid-non-covid}).  This performance is comparable to that of adult CXRs at 98.25\% (Table \ref{tab:covid-non-covid}) in the RSNA \cite{rsna} and RICORD \cite{ricord} test set. This is a strong indicator that the pre-training has helped prevent large-scale overfitting onto small data and adapt to data distribution shifts during the fine-tuning step.

For qualitative analysis and model interpretability, the attention heat maps of the models were examined by an experienced pulmonologist. In most cases, it was found that the model's attention is attending to the regions of interest. However, in some cases, the model attention was also observed in regions outside the lung. Hence, in the future, further analysis for more quantifiable model interpretability is required. However, the quantitative results, specifically on the pediatric data show the effectiveness of the SS-CXR framework in the lung disease classification tasks.   

\section{Conclusions}
\footnotetext[3]{estimated from sensitivity, specificity, positive and negative cases}
Chest x-rays are among the most widely used imaging modalities in radiology. Since the pandemic, CXRs have been used for the diagnosis and prognosis of covid-19, particularly for patients presented to the emergency department. Given the significance of CXR imaging in this context and the increased stress on healthcare facilities, we proposed a paradigm where self-supervised pre-training is employed to solve domain-specific downstream tasks related to CXRs, including classification and segmentation. We demonstrated the merit of the proposed approach on multi-class classification and semantic segmentation tasks, achieving state-of-the-art performance. The performance improvements achieved by the advocated self-learning method were particularly notable in challenging pediatric data.
We showed that the trained model learns useful representations (from adult data) that were transferable to the pediatric data. We are the first to show the application of vision transformer models to such small data ($<200$ images).  In the future, we intend to extend this to a foundation model for chest radiography by leveraging more data (including computed tomography) during pre-training and downstream tasks for detecting multiple lung diseases such as tuberculosis and nodules.   




{\small
\bibliographystyle{unsrt}
\bibliography{PaperForReview}
}
\newpage
\onecolumn
\appendix

\section{Dataset Description}

\subsection{Data used for Classification Experiments}\label{section:dataset-clf}
The data from CXR-3 is divided into the train and test sets as presented in Table \ref{tab:classification}.  
\begin{table*}[ht]
\setlength\tabcolsep{1pt}
\caption{\textbf{Classification Data Overview}. Table showing the counts of CXR images in each of the datasets that have been used for the classification-related tasks in this paper. The table also shows the various stages for which the dataset has been used like pre-training, fine-tuning, and testing. }
\label{tab:classification}
\begin{center}
{\small %
\begin{tabular}{|l|c|c|c|c|c|c|c|}
\hline 
\multicolumn{8}{|c|}{\textbf{Training Data}}                              \\ \hline \hline
 & \textbf{Total} & \textbf{Healthy} & \textbf{Pneumonia} & \textbf{Covid-19} & \textbf{Pre-training} & \textbf{Fine Tuning} & \textbf{Testing} \\ \hline \hline

\textbf{Fig 1}       & 24    & 0    & 0    & 24    & \cmark & \cmark & \xmark  \\
\textbf{SIRM}       & 943   & 0    & 0    & 943   & \cmark & \cmark & \xmark  \\
\textbf{RICORD}     & 896   & 0    & 0    & 896   & \cmark & \cmark & \xmark  \\
\textbf{RSNA}       & 13588 & 8085 & 5503 & 0     & \cmark & \cmark & \xmark  \\
\textbf{STONYBROOK} & 13636 & 0    & 0    & 13636 & \cmark & \cmark & \xmark  \\
\textbf{BIMCV}      & 200   & 0    & 0    & 200   & \cmark & \cmark & \xmark  \\ \hline \hline

\multicolumn{8}{|c|}{\textbf{Unseen Data Distribution}}                              \\ \hline \hline
\textbf{COHEN}      & 674   & 404  & 0    & 270   & \xmark  & \xmark  & \cmark \\
\textbf{ACTMED}     & 132   & 107  & 0    & 25    & \xmark  & \xmark  & \cmark  \\ \hline

\multicolumn{8}{|c|}{\textbf{Test Data (Training Data Distribution)}}                              \\ \hline \hline
\textbf{RICORD}     & 200   & 0    & 0    & 200   & \xmark  & \xmark  & \cmark \\
\textbf{RSNA}       & 200   & 100  & 100  & 0     & \xmark  & \xmark  & \cmark \\ \hline \hline
\multicolumn{8}{|c|}{\textbf{Pediatric Data (Shifted Data Distribution)}}                            \\ \hline
\textbf{Pediatric}           & 189   & 90  & 0    & 99   & \xmark  & \cmark  & \cmark  \\ \hline
\end{tabular}
}
\end{center}
\end{table*}

We used the train data (in the \textit{train.txt} file from Kaggle) for the training and report evaluation metrics for the test set (in the \textit{test.txt} file from Kaggle). Further, to simulate the real-world use case where a different site or location may not be available for training, we removed the ACTMED\cite{actmed} and COHEN\cite{cohen} data from the pre-training and fine-tuning stages and used them for model testing. The test data from RICORD and RSNA site is called the training data distribution, since both these sites are part of the training data used for pre-training and fine-tuning.

\subsection{Pediatric Data}\label{section:cnhdataset}
\begin{figure*}[ht]
\begin{center}  
\includegraphics[width=0.7\textwidth]{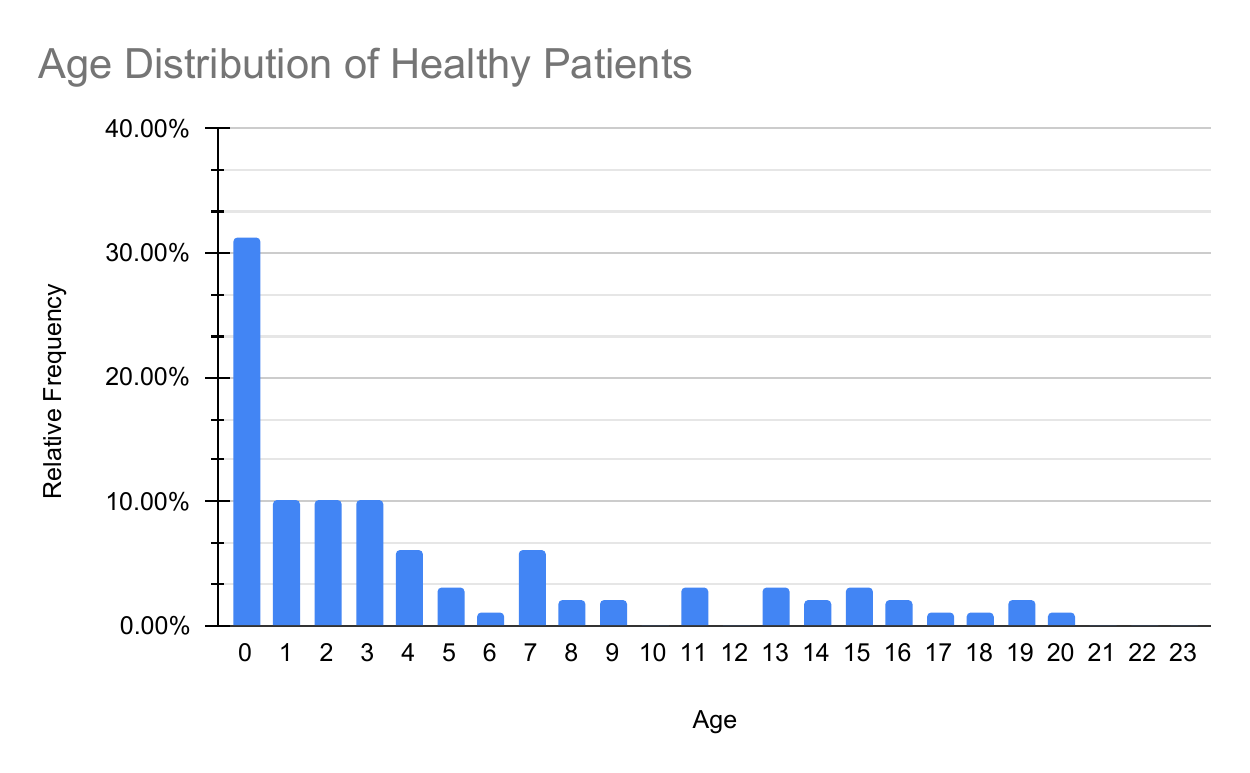}
\end{center}
\caption{\textbf{Age Distribution of Healthy Pediatric Data}. The frequency distribution of the patient age for healthy patients was collected as part of the pediatric dataset.}
\label{fig:healthydata}
\end{figure*}

This dataset contains 90 healthy samples as can be seen in Table \ref{tab:classification}. This subset was obtained by going through reports of each and every patient in the PACS system and looking for the terms "clear lung", "fever", "cough", "wheezing" and "asthma" in the accompanying report as suggested by a senior pediatric pulmonary physician. A filter was applied to get samples that have antero-posterior (AP)/postero-anterior (PA) views to be consistent with the CXR-3 dataset. All patients were in the age range of 0 to 21 years of age with most of them between 0 to 1 year. A detailed age distribution is shown in Figure \ref{fig:healthydata}.

\begin{figure*}[ht]
\begin{center}  
\includegraphics[width=0.7\textwidth]{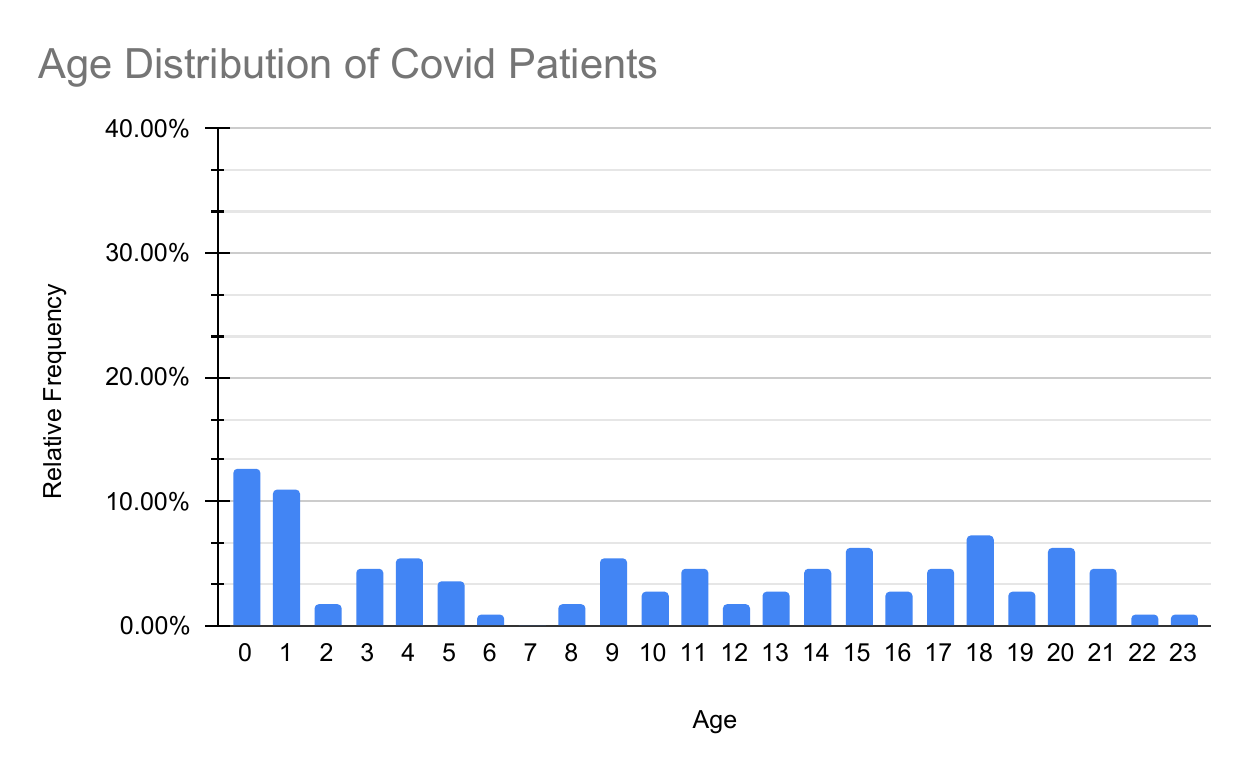}
\end{center}
\caption{\textbf{Age Distribution of Healthy Pediatric Data}. The frequency distribution of the patient age for the covid-19 patients was collected as part of the pediatric dataset.}
\label{fig:coviddata}
\end{figure*}

The pediatric covid-19 positive cases (99) represent patients who visited the ED of the hospital and had a positive RT-PCR anytime before discharge. Furthermore, a filter was applied to look for patients where at least one CXR and an electronic medical report was available. All patients were in an age range of 0 to 23 years with most of them between 0 to 2 years. The detailed age distribution is shown in Figure \ref{fig:coviddata}. The pediatric CXR data were curated in-house after approval from the Internal Review Board (IRB) at the hospital.

\clearpage
\subsection{Segmentation Data}\label{section:dataset-seg}

For the segmentation tasks, none of the data were used in the pre-training stage. The fine-tuning for the downstream segmentation task was performed on data from Shenzhen \cite{jaeger2013automatic} and Belarus sites. The evaluation metrics for the segmentation task are reported on the Montgomery County dataset \cite{candemir2013lung}.

\begin{table}[ht]
 \caption{\textbf{Segmentation Data Overview}. Table showing the counts of CXR images in each of the datasets that have been used for the segmentation-related tasks in the paper. The table shows the various stages for which the data have been used including pre-training, fine-tuning, and testing. The pre-training data is the same as from Table \ref{tab:classification}.}
\label{tab:segmentation}
\begin{center}
\begin{tabular}{|l|c|c|c|c|}
\hline
& \textbf{Total CXRs} & \textbf{Pre-training} & \textbf{Fine Tuning} & \textbf{Testing} \\ \hline \hline
\textbf{Shenzhen}    & 566            & \xmark                   & \cmark                 & \xmark               \\
\textbf{Belarus}     & 97             & \xmark                   & \cmark                & \xmark    \\ \hline

\textbf{Montgomery} & 138            & \xmark                   & \xmark                  & \cmark             \\ \hline

\end{tabular}
\end{center}
\end{table}

\section{Segmentation Results}\label{section:appendix-seg}

This section presents additional visual results for the lung segmentation performance of various machine learning models used in this study. In particular, we present a comparison of our proposed methodology with benchmark (UNETR) segmentation and ImageNet-based pre-training. 

\begin{figure*}[ht]
\begin{center}  
\includegraphics[width=1\textwidth]{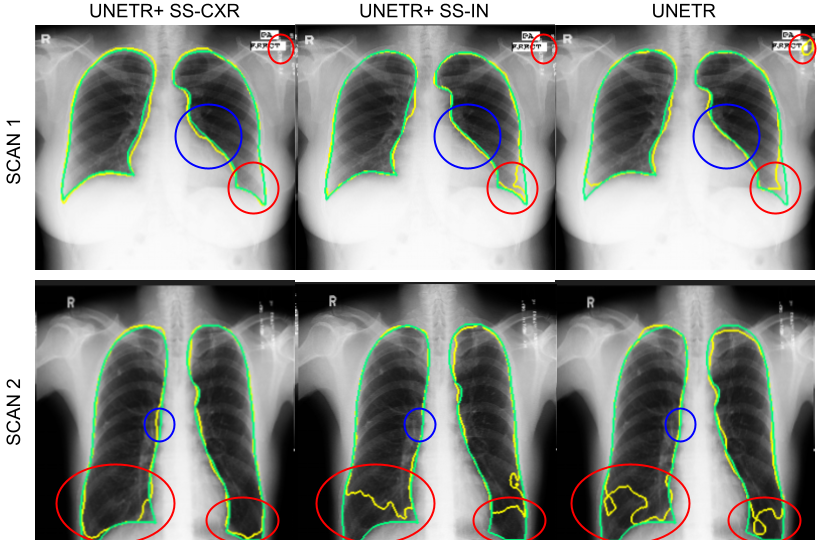}
\end{center}
\caption{\textbf{Segmentation of Lungs}. The yellow polygons are the segmentation predictions for two sample scans from the test set (Montgomery dataset \cite{candemir2013lung}) and the green polygon traces the outline of the ground-truth. The red and blue circles highlight the areas where there is a significant difference in predictions from different models. }
\label{fig:segmentation-zoomed}
\end{figure*}

\begin{figure*}[ht]
\begin{center}  
\includegraphics[width=1\textwidth]{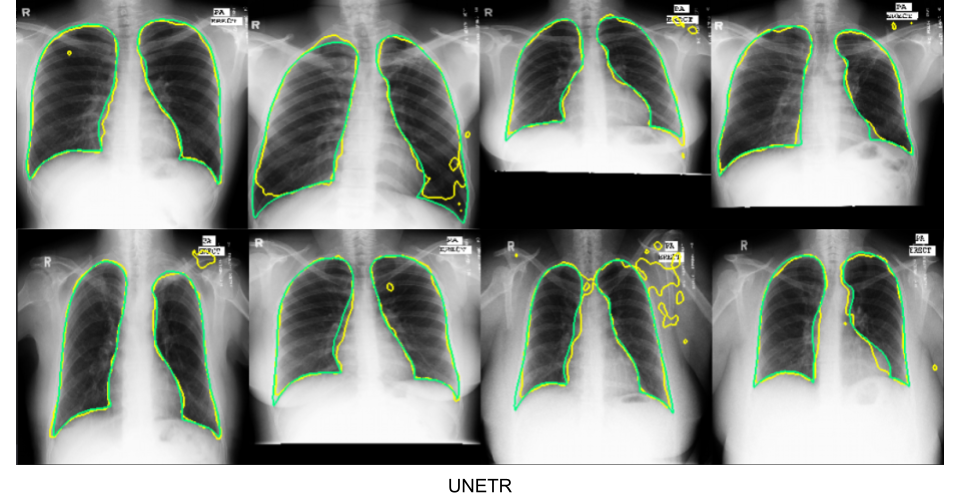}
\end{center}
\caption{\textbf{Segmentation of lungs using UNETR\cite{unetr}}. The yellow polygons are predicted segmentation for sample scans from the test set (Montgomery dataset \cite{candemir2013lung}) and the green polygon traces the outline of the ground-truth.}
\end{figure*}

\begin{figure*}[ht]
\begin{center}  
\includegraphics[width=1\textwidth]{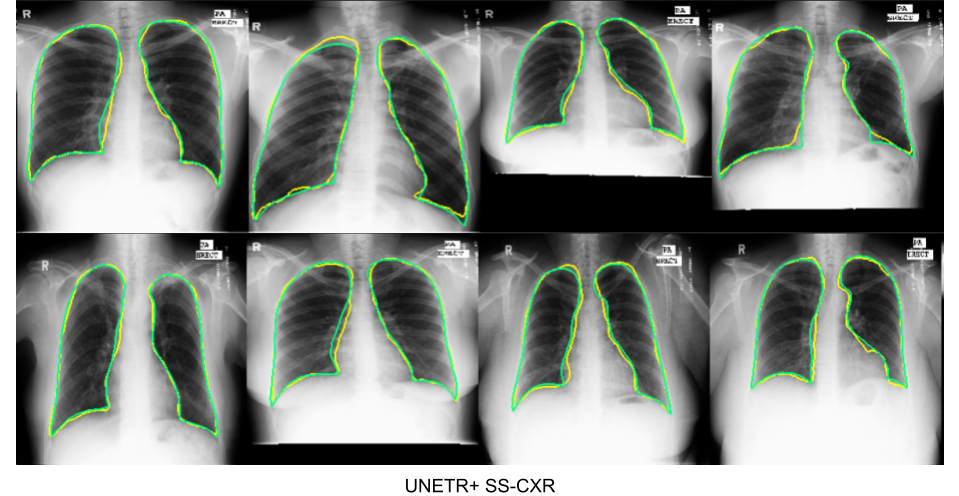}
\end{center}
\caption{\textbf{Segmentation of lungs using UNETR with SS-CXR}. The yellow polygons are predicted segmentation for sample scans from the test set (Montgomery dataset \cite{candemir2013lung}) and the green polygon traces the outline of the ground-truth.}
\end{figure*}

\begin{figure*}[ht]
\begin{center}  
\includegraphics[width=1\linewidth]{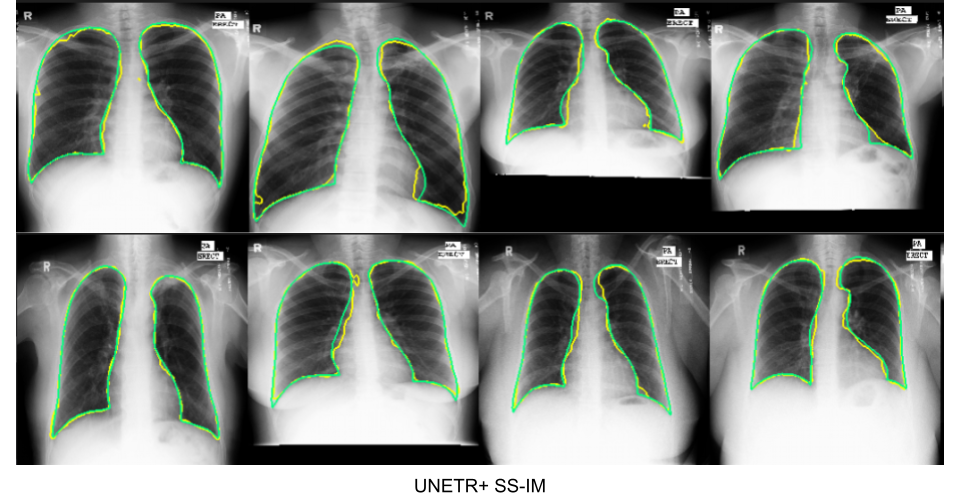}
\end{center}
\caption{\textbf{Segmentation of lungs using UNETR with SS-IM}. The yellow polygons are predicted segmentation for sample scans from the test set (Montgomery dataset \cite{candemir2013lung}) and the green polygon traces the outline of the ground-truth.}
\end{figure*}
\clearpage
\section{PR and ROC Curves with Corresponding AUC}\label{section:appendix-roc1}
\begin{figure*}[ht]
\begin{center}  
\includegraphics[width=1\textwidth]{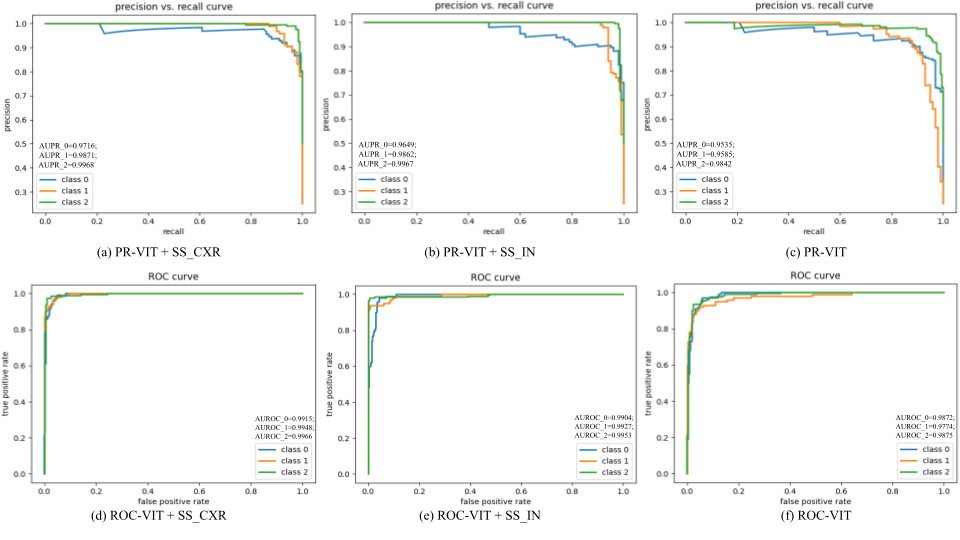}
\end{center}
\caption{\textbf{Receiver Operating Characteristics (ROC) and Precision-Recall (PR) Curves.} Figures a, b, and c show the PR curves of ViT+SS-CXR, ViT+SS-IN and ViT respectively in a one-vs-all classification scenario for multi-class prediction. Figures c, d, and e show the ROC curves in a similar scenario. Class 0 corresponds to healthy vs Pneumonuia \& Covid-19, Class 1 corresponds to Pneumonia vs Healthy \& Covid-19 and Class 2 corresponds to Covid-19 vs Healthy \& Pneumonia.}
\label{fig:plotcurves1}
\end{figure*}

\begin{figure*}[ht]
\begin{center}  
\includegraphics[width=1\textwidth]{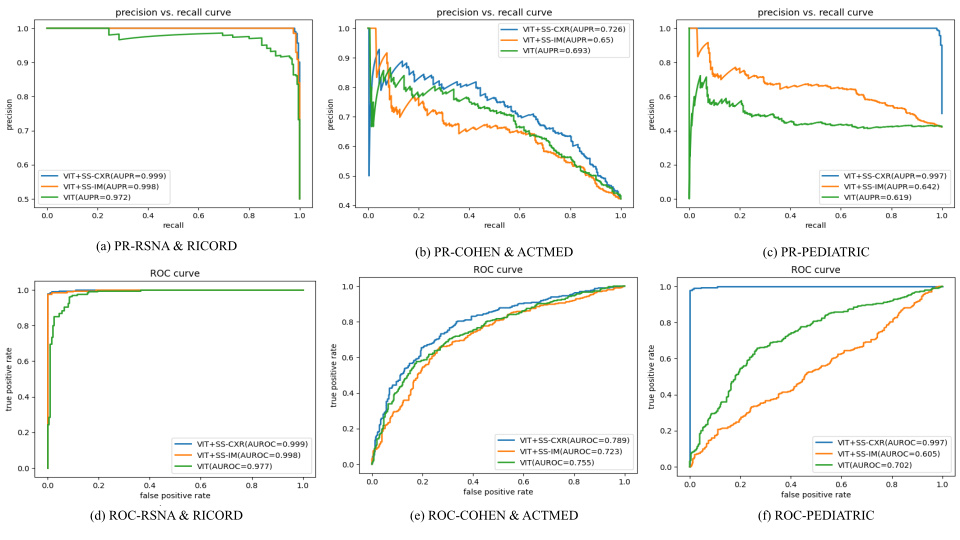}
\end{center}
\caption{PR (a, b, and c) and ROC (d, e, and f) curves for covid/non-covid classification using RSNA \& Ricord test set (a \& d), unseen data distribution (Cohen \& Actmed) (b \& e), and shifted data distribution (pediatric) (c \& f) corresponding to classification results presented in Table 1. }
\label{fig:plotcurves2}
\end{figure*}

\begin{figure*}[ht]
\begin{center}  
\includegraphics[width=1\textwidth]{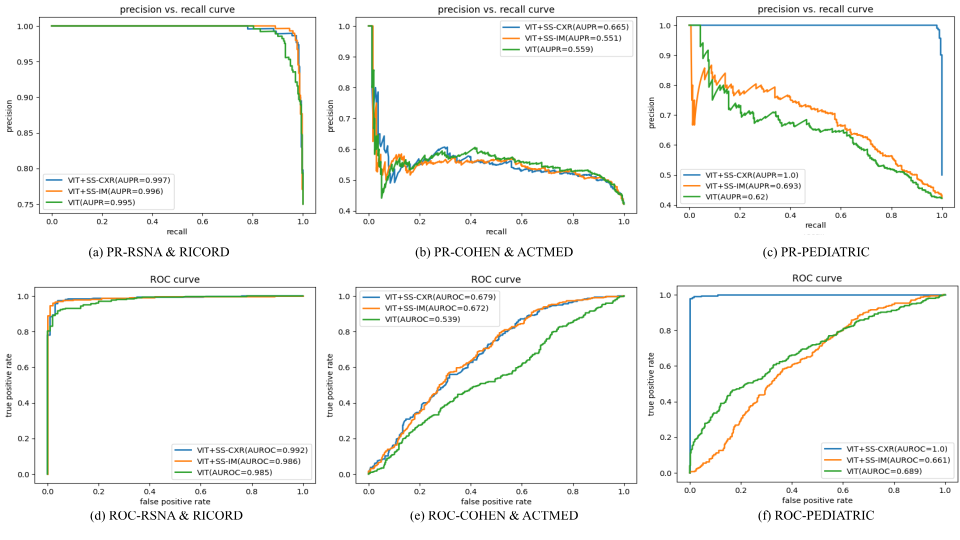}
\end{center}
\caption{PR (a, b, and c) and ROC (d, e, and f) curves for healthy/pathology classification using RSNA \& Ricord test set (a \& d), unseen data distribution (Cohen \& Actmed) (b \& d), and shifted data distribution (pediatric) (c \& f) corresponding to classification results presented in Table 2.}
\label{fig:plotcurves3}
\end{figure*}

\begin{figure*}[ht]
\begin{center}  
\includegraphics[width=0.9\textwidth]{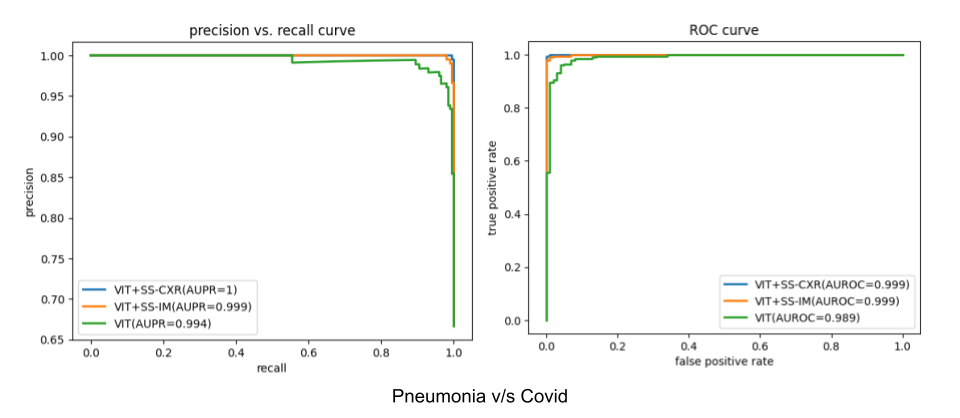}
\end{center}
\caption{PR and ROC curves for binary classification of Pneumonia and covid-19 for the test set (samples from RSNA \& RICORD sites) corresponding to results presented in Table 3.}
\label{fig:segmentation2}
\end{figure*}

\begin{figure*}[ht]
\begin{center}  
\includegraphics[width=0.95\textwidth]{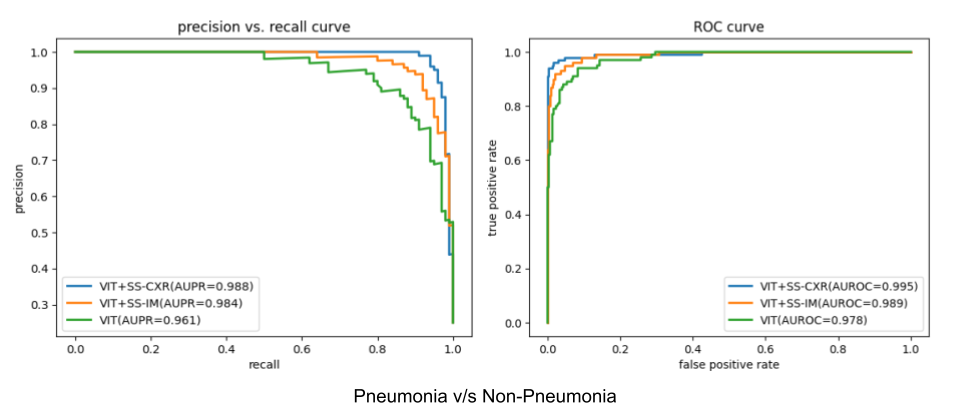}
\end{center}
\caption{PR and ROC curves for binary classification of Pneumonia and Non-Pneumonia for the test set (samples from RSNA \& RICORD sites) corresponding to results presented in Table 3.}
\label{fig:segmentation3}
\end{figure*}

\clearpage
\section{GMML Augmentation Showcase}\label{section:appendix-roc}
\begin{figure*}[ht]
    \centering
    \includegraphics[width=1\linewidth]{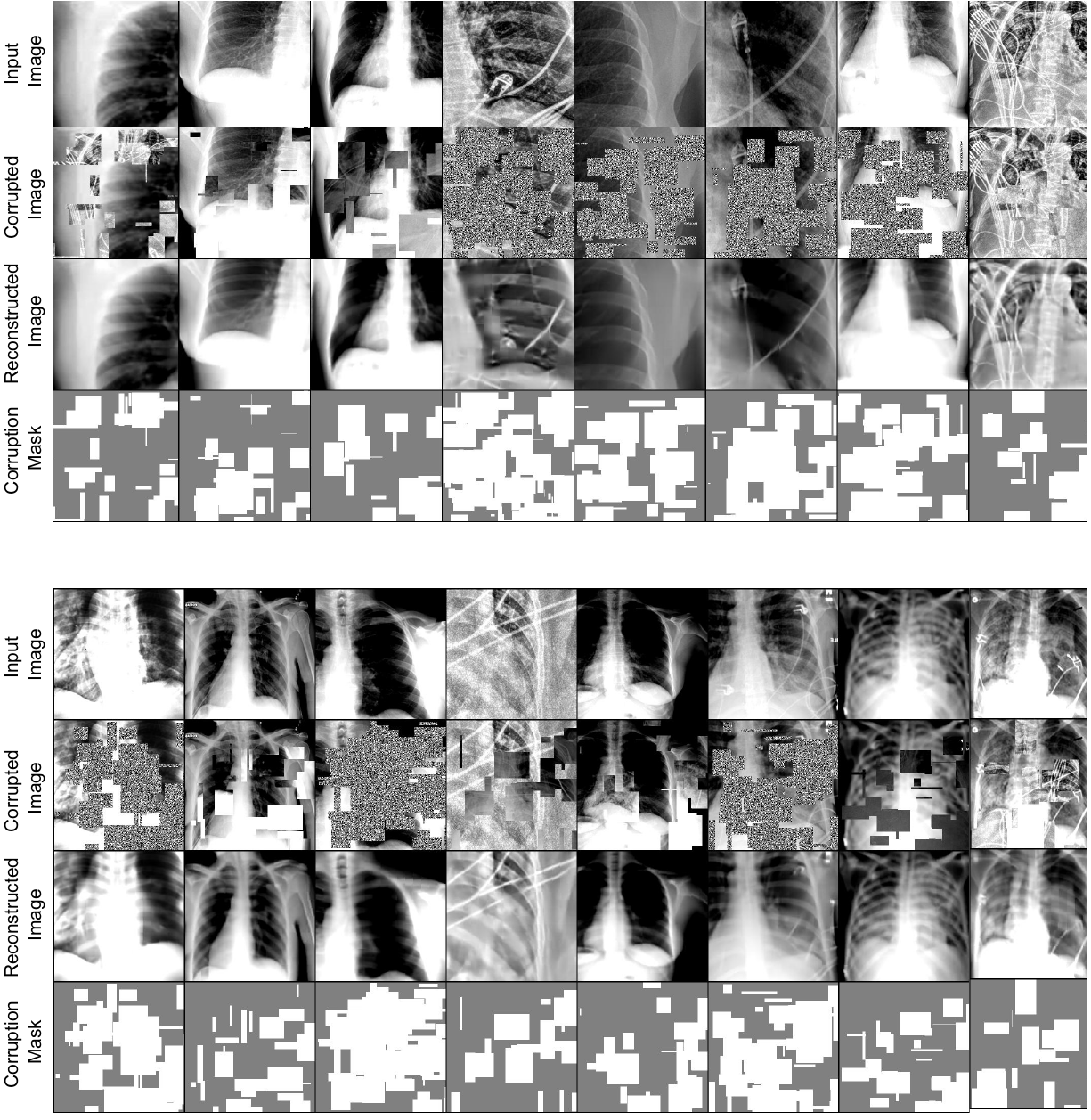}
    \caption{Some examples showing the GMML CXR image manipulation including the input image, corrupted image, reconstructed image, and the corruption mask.}
    \label{fig:GMMLmani}
\end{figure*}

\clearpage
\section{CXR Foundation Model Strategy}\label{section:appendix-roc2}
In the medical imaging domain, generally there is a lack of pre-trained models that are publicly available due to concerns associated with patient privacy. Also in clinical setting there is a lack of access to compute power for large scale pre-training on datasets having a significant size. To overcome these challenges, we intend to make the pre-trained weights of the vision transformer available for both SS-CXR and SS-IM paradigms. The aim is to allow the research community in clinical settings to use these pre-trained weights for a downstream task specific fine-tuning utilizing the learned representation from CXR data. Hence, lay the ground work for the vision transformer-based foundation model for CXR tasks. To achieve this goal, we intend to use the MoNAI framework (\url{monai.io}), which is an open source and widely used collaborative framework  built for accelerating research and clinical collaboration in the medical imaging domain.






\end{document}